\documentclass[10pt,a4paper,notitlepage]{iopart}

\usepackage{iopams} 
\expandafter\let\csname equation*\endcsname\relax
\expandafter\let\csname endequation*\endcsname\relax
\usepackage{amsmath} 
\usepackage{mathtools} 

\usepackage{enumitem}
\usepackage{graphicx}
\usepackage{epstopdf}
\usepackage{nicefrac}
\usepackage[caption=false]{subfig}

\usepackage[bookmarksopen=true,bookmarksopenlevel=\maxdimen, bookmarksnumbered=true]{hyperref}
\usepackage[all]{hypcap}

\usepackage[usenames, dvipsnames]{color}

\begin{document}

\title{Density distributions and depth in flocks}

\author{J M Lewis$^1$ and M S Turner$^{1,2}$}

\address{$^1$ Centre for Complexity Science, University of Warwick, Coventry CV4 7AL, UK}
\address{$^2$ Department of Physics, University of Warwick, Coventry CV4 7AL, UK}

\ead{M.S.Turner@warwick.ac.uk}


\begin{abstract}  
Recent experimental evidence suggests that interactions in flocks of birds do not involve a characteristic length scale. Bird flocks have also been revealed to have an inhomogeneous density distribution, with the density of birds near the border greater than near the centre. We introduce a strictly metric-free model for collective behaviour that incorporates a distributed motional bias, providing control of the density distribution. A simple version of this model is then able to provide a good fit to published data for the density variation across flocks of Starlings. We find that it is necessary for individuals on the edge of the flock to have an inward motional bias but that birds in the interior of the flock instead must have an outward bias. We discuss the ability of individuals to determine their depth within a flock and show how this might be achieved by relatively simple analysis of their visual environment.
\end{abstract}



\vspace{10pt} \begin{indented} \item[]\textrm Keywords: swarming, topological interaction, metric-free, distributed motional bias, inhomogeneous swarm density, flock depth, self-organisation.\par \end{indented}

\vspace{28pt} 

\begin{indented} 
	\item[] \textrm This Accepted Manuscript is available for reuse under a CC BY-NC-ND 3.0 licence after the 12 month embargo period provided that all the terms of the licence are adhered to.
	\item[] \textrm This is a peer-reviewed, un-copyedited version of an article accepted for publication/published in Journal of Physics D: Applied Physics. IOP Publishing Ltd is not responsible for any errors or omissions in this version of the manuscript or any version derived from it. The Version of Record is available online at \href{https://doi.org/10.1088/1361-6463/aa942f}{10.1088/1361-6463/aa942f}.
\end{indented}

\ioptwocol 


\section{Introduction} \label{sec:introduction}

Swarming is the collective behaviour of animal aggregations, and can be observed in the flocking of birds \cite{king2012murmurations,emlen1952flocking,feare1984starling}, fish shoaling \cite{parrish1999complexity,pitcher1983heuristic,pitcher1986functions}, mammal herding \cite{gueron1996dynamics,couzin2003self}, and insect swarming \cite{buhl2006disorder,okubo1986dynamical}. Human crowds can also display this sort of collective trait \cite{helbing1995social,helbing2001self}. The emergence of global orientational order in groups of moving animals is arguably the most striking consequence of this type of social behaviour \cite{okubo1986dynamical,sumpter2006principles}. In these systems collective animal behaviour is not thought to arise from centralised coordination but rather the system is believed to exhibit self-organisation due to the local rules of the interacting elements. This results in coherent motion with local rules manifesting global order \cite{camazine2003self,nicolis1977self}.

In recent years, a large number of theoretical models have been developed in which local interaction rules give rise to global ordering in animal systems \cite{giardina2008collective,carrillo2014derivation,carrillo2017review} however empirical studies have been more rare \cite{lukeman2010inferring,katz2011inferring}. Testing models against data is essential if we are to determine which sorts of model give rise to specific characteristics: many models can generate some form of swarming, but which of these models give rise to swarms that resemble those seen in nature? It has been suggested that the specific interaction mechanism may vary with species and for some systems an interaction based on neighbour distance appears to be a good fit \cite{vicsek2012collective}. In contrast, recent field studies have reconstructed the internal dynamics of large flocks of Starlings and have determined that their nearest-neighbour interactions do not depend on interaction range \cite{ballerini2008empirical,ballerini2008interaction,cavagna2010scale}. 

Developing models with this \textit{metric-free} characteristic is technically challenging as they typically support a zero density steady-state, such as described in the work of Ginelli and Chat\'{e} \cite{ginelli2010relevance} in which diffusive expansion continues indefinitely. Pearce and Turner \cite{pearce2014density} describe a model that regulates swarm density using a motional bias on surface individuals and \textit{topological} interaction rules, preserving the metric-free nature of the model and also generating a steady-state with finite spatial extent. This Strictly Metric-Free (SMF) model is therefore useful to compare with observations of bird flocks as it can produce bounded swarms in open boundary conditions. However we will show that, in its simplest form, it yields density distributions that are rather different to those observed.

In this work we propose a fully topological (metric-free) 3-dimensional model which includes a motional bias that is tunable throughout the swarm and not just on its surface. This bias has a topological character, preserving the fully topological nature of the model. Our aim is to explore the regulation of density across flocks of birds. We are motivated by findings from a field study \cite{ballerini2008empirical} that reports a nonhomogeneous density variation across flocks of Starlings, specifically a higher density at the border of the flock than in the centre. This observation is counter to what has been observed in some other models of collective behaviour \cite{kunz2003artificial}. It is also counter-intuitive in relation to some theories of animal behaviour, such as the selfish herd hypothesis \cite{hamilton1971geometry} in which the centre of the group would be the safest location and all individuals might therefore be expected to seek to occupy it. We show that our metric-free distributed motional bias model is able to support behaviour consistent with these empirical observations.

The model is introduced in section \ref{sec:model}. The methods used to measure aggregate densities and fit the model to data are described in section \ref{sec:methodology}. The resultant model and swarm density profiles are presented in section \ref{sec:results}. Additionally, a biologically motivated basis for an individual determining their depth from within the flock is presented and discussed in section \ref{sec:topo-depth}. Concluding remarks are in section \ref{sec:conclusion}.


\section{Description of the model} \label{sec:model}

The model we propose begins with the surface bounding effect introduced in the SMF model \cite{pearce2014density} and extends it to act on all individuals in the aggregate with strength prescribed by a function of the topological depth of the individual within the swarm. In contrast to classic models of self-propelled particles, such as those by Vicsek \textit{et al.} \cite{vicsek1995novel}, we identify two particles to be neighbours if they are directly connected to each other under a Voronoi tessellation \cite{lee1980two,okabe2009spatial}. This is constructed for the particle positions at each time step, thus defining interacting neighbours as those in neighbouring Voronoi cells (i.e. particles which share an edge in the Delaunay triangulation of all particle locations).

We use this tessellation to determine \textit{topological depth} for each of the particles in the dynamic aggregate (flock). We first identify a shell, or set, of particles as being those that occupy an infinite Voronoi cell. These are denoted as occupying shell $0$ and correspond to particles that are on the convex hull of the system \cite{okabe2009spatial}. Particles that are connected to these shell $0$ particles via Delaunay edges, but that are not themselves members of shell $0$, are defined to lie in shell $1$. This process is repeated iteratively until all particles are assigned a shell number. This labelling encodes topological depth as it relates to the shortest path length from the border through the graph defined via the Delaunay triangulation. A driving term can then be included in the equation of motion that provides a motional bias on each particle. The direction of this bias (loosely ``inwards'' or ``outwards''), is derived using the locations of its neighbours on the same shell. 

\begin{figure} []
	\capstart
	\includegraphics[width=.48\textwidth]{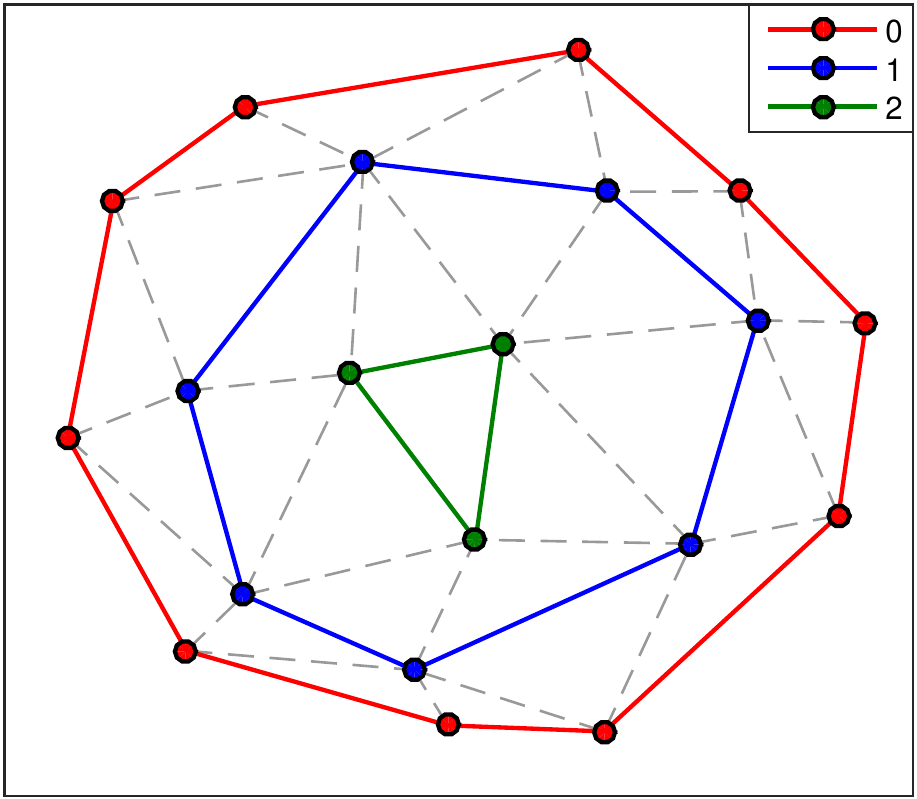}
	\caption{Schematic of system topology (shown as a 2D sketch for clarity). Particles (circles) are nodes of a Delaunay graph obtained via Voronoi tessellation. If a particle $i$ shares any edge (coloured or dashed) with particle $j$, then $j$ is in the set of neighbours $B_i$. Coalignment involves reorientation towards the average orientation of these particles $\langle\underline{\hat{v}}_j^t\rangle_{j\in B_i}$. Voronoi shells are denoted by coloured edges; red: 0, blue: 1, green: 2. The set of all particles with same shell number $\kappa$ is $C_\kappa$, with the shell index of particle $i$ denoted $\kappa(i)$. If a neighbour to particle $i$ is also on the same shell $\kappa(i)$ (i.e. shares a coloured, not dashed, edge), then it is in set of shell neighbours $S_i=B_i\cap C_{\kappa(i)}$. The motional bias acts in the direction of the average of unit vectors pointing from particle $i$ to each of the shell neighbours $\langle\hat{\underline{r}}_{ij}^t \rangle_{j \in S_i}$. An illustrative construction of the bounding term for each particle is provided in the Supplemental Materials. This two dimensional construction naturally extends to the three dimensional model discussed in the text (colour online).}
	\label{fig:model_schematic}
\end{figure}

The interaction rules governing all $N$ identical particles in the system are shown in equations \ref{eq:dmbsmf1}--\ref{eq:dmbsmf3}. 
\begin{align} 
&\underline{r}_i^{t+1} = \underline{r}_i^t + v_0 \underline{\hat{v}}_i^t \label{eq:dmbsmf1} \\
&\underline{v}_i^{t+1} = (1-\phi_n) \underline{\hat{\mu}}_i^t + \phi_n \underline{\hat{\eta}}_i^t \label{eq:dmbsmf2} \\
&\underline{\mu}_i^t = f_i \langle\underline{\hat{r}}_{ij}^t\rangle_{j\in S_i} + (1-f_i) \vartheta\big(\langle\underline{\hat{v}}_j^t\rangle_{j\in B_i}\big) \label{eq:dmbsmf3}
\end{align}
They involve the position $\underline{r}_i^{t}$ of particle $i$ at discrete time $t$, having direction of motion $\underline{\hat{v}}_i^t$ and constant speed $v_0$, which is set equal to unity in what follows. The ``hat'' symbol $\hat{\;\;}$ denotes a normalised (unit) vector and angled brackets $\langle\cdots\rangle$ indicate an average over the indicated particle subset. The operator $\vartheta(\;)$ performs normalisation via $\vartheta(\underline{w})=\nicefrac{\underline{w}}{|\underline{w}|}$ and $\underline{\hat{r}}_{ij}^t$ denotes the unit vector pointing from particle $i$ to particle $j$ at discrete time step $t$. The parameter $\phi_n$ encodes the strength of the (vectorial \cite{chate2008collective}) noise applied to each particle, multiplied by a random unit vector obeying $\langle \underline{\hat{\eta}}_i^t\rangle=0$ and $\langle \underline{\hat{\eta}}_i^t\cdot\underline{\hat{\eta}}_j^{t'}\rangle=\delta_{i,j}\delta_{t,t'}$. The neighbours of particle $i$ are denoted $B_i$ and particles which share the same shell number $\kappa(i)$ as particle $i$ form the set $C_{\kappa(i)}$. Therefore we denote the set of shell neighbours of $i$ as the intersection $S_i=B_i\cap C_{\kappa(i)}$. We average over the unit vectors pointing from particle $i$ to members of this set. In addition, figure \ref{fig:model_schematic} shows how shell 0 is defined as the members on the convex hull of the system, and also the procedure for identifying all other shells. Equation~(\ref{eq:dmbsmf1}) represents a simple vectorial particle translation along the current velocity. Equation~(\ref{eq:dmbsmf2}) encodes an update rule for the velocity that includes both some deterministic driving terms, weight $(1- \phi_n)$, and some stochastic noise, weight $\phi_n$. Thus $\phi_n$, the degree of noise, is an important control parameter in what follows. Equation \ref{eq:dmbsmf3} defines the deterministic driving terms. It is comprised of two terms, the first, with weight $f_i$, encodes the motional bias constructed from the shell geometry, as described and the second term, with weight $(1 - f_i)$, provides co-alignment of each particle with its neighbours.

We denote $f_i$ as the ``bounding function'', which encodes the relative strength of the bounding effect on each Voronoi shell. Changing this allows us to tune the bounding of the model across the aggregation as we wish. If we choose $f_i$ to have the form of equation \ref{eq:dmbsmf4}, where $\phi_e$ is a parameter controlling the strength of the border shell effect, then we can recover the Strictly Metric-Free (SMF) model of Pearce and Turner \cite{pearce2014density} in its entirety. If instead we choose $f_i=0 \;\,\forall i$ then we recover the unbounded metric-free model of Ginelli and Chat\'{e} \cite{ginelli2010relevance}.
\begin{align} \label{eq:dmbsmf4} 
f_{i}  =
\begin{cases}
\; \phi_e & \underline{r}_i^t \in C_0 \\
\; 0 & \text{otherwise}
\end{cases}
\end{align}

In our model, which uses topological shell depth, the value of $f_i$ is the same for all particles in the same shell and can therefore be mapped to a lower dimensional parameter set $f_{\kappa(i)}$. We believe that this generalisation of the SMF model is natural, allowing us to describe the motional bias, not as a specific characteristic for a subset of birds, but as a rule for all birds that has a strength that depends on the relative depth of an individual in the swarm. 

\section{Methodology} \label{sec:methodology}

We are interested in measuring the density variation across our simulated swarms. As we wish to compare directly to the empirical study of Starling murmurations \cite{ballerini2008empirical} we seek to compute this in a similar fashion. The type of flocks which were studied in \cite{ballerini2008empirical} were non-columnar and compact, with sharp borders, containing on the order of hundreds to thousands of birds, and which moved nearly linearly for sufficiently long times so as to treat their behaviour as near steady-state. The type of density variation we are interested in here is the density profile across flocks in this steady state, which is observed to be higher near the edge and to decrease toward the centre: It is not the propagating density waves observed in response to specific events, such as turning or shock.

We determine the spatial extent of simulated swarms using the $\alpha$-shape method \cite{ballerini2008empirical,edelsbrunner1994three}, which allows for the presence of concavities within the swarm to the scale of $\alpha$. To measure density, individuals with distance less than $\delta$ from the border were removed and a new border of the reduced flock was computed. The reduced density was computed using this reduced volume and the number of internal birds. This process was repeated until the flock was empty (i.e.\ less than four members remaining such that no tetrahedra, and hence no volume, can be determined). 

Simulated swarms typically have a non-negligible degree of concavity (as is also observed in the empirical study), therefore allowing for presence of a non-convex border is natural. Fixing the convexity scale $\alpha$ is non-trivial as we are not dealing with a few observations, but thousands of configurational snapshots per simulation, therefore we cannot do this manually (as is described in \cite{cavagna2008starflag}). Instead we obtain a sensible estimate for $\alpha$ by selecting the smallest value possible that leaves the particle aggregation as a single connected component. This fixes the convexity scale throughout. We must also make a choice of the flock reduction parameter $\delta$ as this impacts on our measurement and ability to compare with the data. We select a value which on average provides a similar number of flock reduction iterations as the field study (which is 7). 

In order to prevent this choice from impacting our measurements we scale the reduction so that shell number is mapped to the domain $[0,1]$ with $0$ corresponding to the first reduction and $1$ the final reduced flock. This also allows for a much easier comparison with the observational data; we can map that data to the same domain and perform cubic splines interpolation to allow query of comparison points between simulated and empirical data. Additionally we normalise the density data such that the first flock density measurement is $1$, which makes our measurements and comparisons dimensionless, and allows us to look primarily at the density gradient across the aggregation. These transformations allow us to compare our simulation data more easily with the empirical data and minimises the impact of possible differences in choice of parameters.

Our primary goal is to identify a bounding function $f_{\kappa}$ that can produce simulations with density profiles that provide a good fit to the empirical data. There is some freedom in how one might parametrise $f_{\kappa}$. We choose $f_{\kappa}$ to be linear in shell depth (parametrised via gradient $a$ and intercept $b$).  We allow the bounding strength on shell 0 individuals to be a separate parameter $\phi_e$ in order to include models in which individuals on the edge behave differently from the bulk.

We then use the Simultaneous Perturbation Stochastic Approximation (SPSA) algorithm \cite{spall2005introduction,spall1992multivariate} for recursive optimisation of bounding function parameters ($\phi_e$, $a$, $b$) using gain sequences with suggested practical values from \cite{spall1998implementation}. We used the mean-squared difference between simulated and empirical data, averaged over a specified number of density evaluations, as the cost function estimate. Using this method allows for a principled stochastic search of the parameter space and can be performed in parallel. Fresh simulations were performed at each parameter update, due to the presence of hysteresis in these types of systems \cite{chate2008collective,gregoire2004onset}.


\section{Results} \label{sec:results}

\begin{figure} []
	\vspace{0px}
	\capstart
	\includegraphics[width=.483\textwidth]{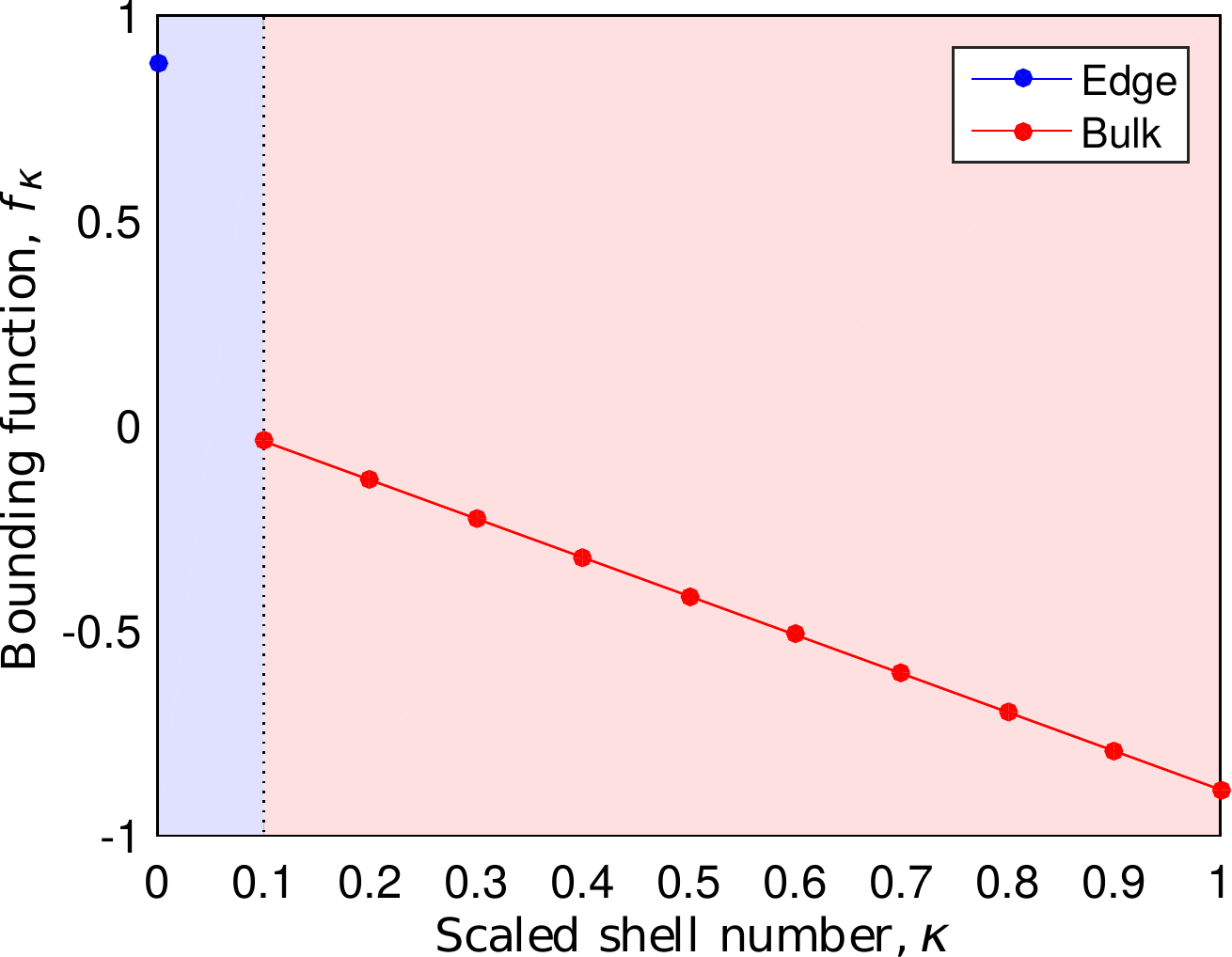}
	\caption{Distribution of motional bias via bounding function $f_{\kappa}$ shown for parameters fit via stochastic optimisation: $\phi_e = 0.883$, $a = -0.944$, $b = 0.056$. Example is shown for $11$ shells: $0$ to $10$. Blue denotes the surface members, shell 0, whose motional bias is determined by $\phi_e$, and red denotes members in the swarm bulk, with scaled shell number $0 < \kappa \leq 1$ and motional bias linearly parameterised by $ a\kappa + b $ (colour online).}
	\label{fig:opt_bounding_func}
\end{figure}

In order to understand how the density across aggregation varies for swarms which interact in a metric-free fashion we generated simulations of our Distributed Motional Bias Strictly Metric-Free (DMBSMF) model, as described above. As we are interested in simulating real-world behaviour we choose the parameters for the model via stochastic optimisation using the previously described method, directly fitting to empirical data, obtaining fit parameters of $\phi_e = 0.883$, $a = -0.944$, $b = 0.056$. These parameters result in a bounding function $f_{\kappa}$ as displayed in figure \ref{fig:opt_bounding_func}. This translates to a strong surface effect generally pointing toward the centre of the flock, however the bulk of the flock has an outward motional bias of increasing strength as one approaches the centre.

In order to simulate a flock that is comparable to that observed in the field study, we note from the motivating empirical study \cite{ballerini2008empirical} that the flock in question contains $1,360$ reconstructed birds. We also note details from a later study \cite{cavagna2010scale}(S.I.) for the flock in question: $1,571$ reconstructed birds with a measured polarisation of $0.96 \pm 0.03$ (i.e.\ observed flocks in high order regime). We therefore chose to simulate $1,500$ birds with noise parameter $\phi_n = 0.22$, yielding a polarisation of $0.931 \pm 0.003$, which is of similar magnitude to the observed flock. In each instance, we performed a simulation for $20,000$ time steps with the first $10,000$ steps discarded for equilibration of the system. The initial condition is a random (isotropic) orientation and a random location, uniformly distributed within a unit cube, for each individual. We measured the density variation across the flock (as described in section \ref{sec:methodology}) every $10$ time steps after equilibration, resulting in $1,000$ measurements per simulation instance, which are then time-averaged. We combine the results from five independent simulation instances, with final values presented as the mean of these quantities and uncertainties corresponding to standard errors.

\begin{figure} []
	\vspace{2px}
	\capstart
	\includegraphics[width=.479\textwidth]{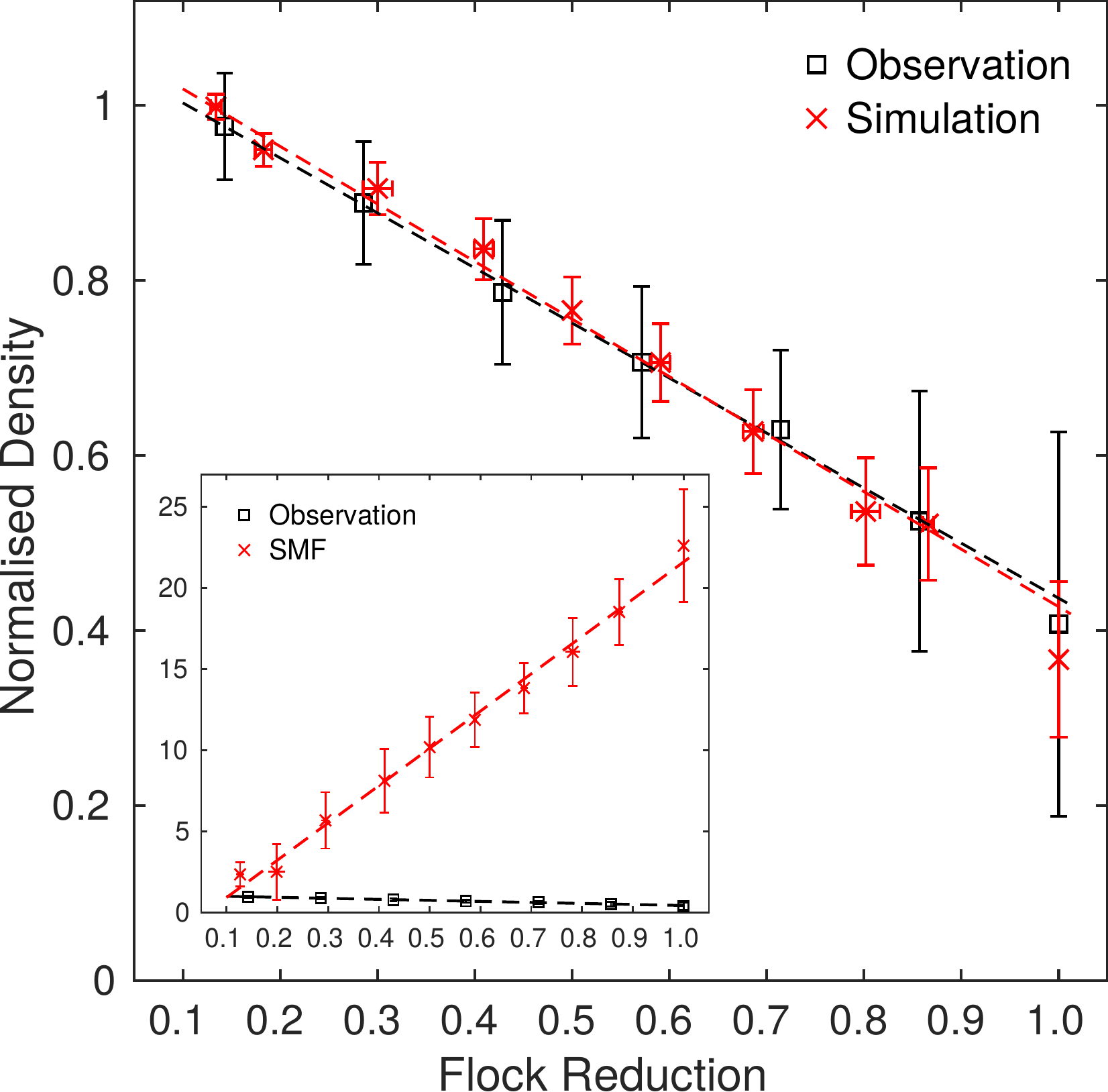}
	\caption{Density variation across aggregation: comparison of empirical data (black squares) reproduced from \cite{ballerini2008empirical} with simulation data from DMBSMF model (red crosses) with parameters $\phi_e = 0.883$, $a = -0.944$, $b = 0.056$ obtained via stochastic optimisation. Simulation data is an average of five time-averaged independent initialisations. Measurements are normalised such that the first flock reduction ($= 0$) has unit density, and a value of $1$ corresponds to the final measurement before a fully reduced (empty) flock. Linear fits show good agreement with the model: flock density is largest on the border and decreases toward the centre at a similar rate. Inset: Similar comparison of empirical data (black squares) with SMF model (red crosses) with parameters $\phi_e = 0.5$, $a = b = 0.0$, as in \cite{pearce2014density}. Axis labels as for main figure. The large disparity between the SMF model and observations highlights the strength of our new DMBSMF model (colour online).}
	\label{fig:opt_vs_data}
\end{figure}

The simulated model matches closely to empirical data of Starling flocks, as can be seen in figure \ref{fig:opt_vs_data}, and produces the observed effect that aggregation density is greater at the border and reduces in what appears to be a linear fashion. The rate of this decrease is also closely matched. This counter-intuitive observation appears to require a model with a surprising motional bias: whilst surface birds move toward the flock centre, ensuring global cohesion, the rest of the flock move toward the border with increasing strength the further from it they are, as determined by topological depth. Naturally then, the number of birds closer to the border of the flock increases and drops off toward the centre due to the strong gradient of the bulk bounding function. 

Our model shares some similarity with another recently proposed flocking model, the ``hybrid projection" model \cite{pearce2014role}, that drives individuals to move towards features in their visual field, specifically the boundaries between light and dark regions, where light/dark encodes the absence/presence of a neighbour in each direction. This model effectively encourages the movement inwards of individuals near the flock border. This is because individuals at the border will experience featureless outward-directed visual fields, resulting in an inward bias. It will also generate a bias outwards from the bulk of the flock as there will typically be more features in the outward-pointing directions than toward the often opaque centre of the flock. It is notable then that the motional bias that fits data from real-world flocks is similar to the effective motional bias present in visual models of this type.

	
\section{Determining topological depth} \label{sec:topo-depth}

A key aspect of our model is the notion of topological depth within the flock. Individuals are assigned a shell number based on this quantity, encoding a non-metric measure of depth as the shortest path length from the individual to a member of the convex hull (shell 0). The motional bias experienced by this individual is a function of shell number, as shown in figure \ref{fig:opt_bounding_func}. It is therefore important to consider the accessibility of this quantity to the individual, from a biological/ sensory perspective - how might flock members determine their shell number? In this section we present a model for how this could be achieved using the degree of anisotropy in an individuals visual field as an indicator of their depth within the flock.

\begin{figure} []
	\centering
	\includegraphics[width=.485\textwidth]{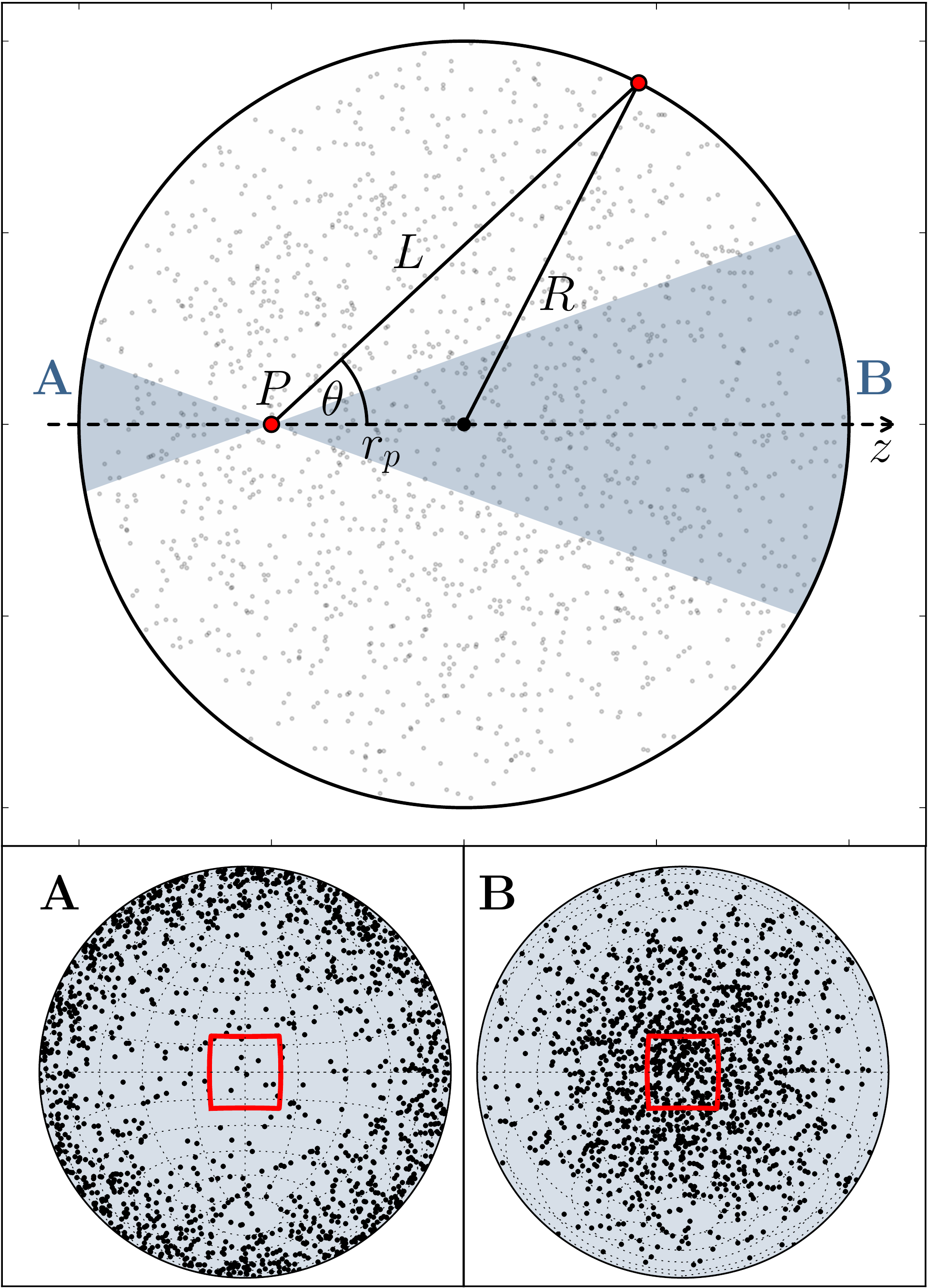}
	\caption{Schematic of simplified description of system: a cross-section of a sphere of radius $R$ with homogeneous mass density $\rho(\underline{r}) = \rho$ for $|\underline{r}| \leq R$ and $0$ otherwise. We consider the visual information available to an individual at point $P$ at a distance $r_p$ from the centre in different directions encoded by $\theta$. $L(\theta$) is the distance from $P$ to the edge of the sphere in the $\theta$ direction. The system is symmetric under rotation about the axis $z$ defined in the direction from $P$ through the centre of the sphere. Areas A and B, shaded in blue, denote an example field of view between $\pm20\deg$ in the negative and positive $z$ direction respectively. In the sub-figures A/B we plot the position of $1500$ flock members relative to $P$, at $r_p = R/2$ from the centre, as Lambert azimuthal equal-area projections centred on the direction of negative/positive $z$ respectively, with the red region denoting bounds of $\pm20\deg$ along each axis. Looking along $z$ through the flock, as in B, one can see a high density of other flock members, however this is drastically reduced when looking in the opposite direction out of the flock, as in A. We use this visual anisotropy as the basis for an individual inferring its depth from deep within the flock.}
	\label{fig:sphere_swarm_schematic}
\end{figure}

We analyse a simplified model of the system in which the density is homogenous, for simplicity. Consider the three-dimensional flock as a sphere $S$ of radius $R$ centred on the origin with particle mass distributed uniformly within this sphere. For a point $P$ on or inside the sphere we can define an axis $z$ along the vector from $P$ to the sphere centre at the origin, as seen in figure \ref{fig:sphere_swarm_schematic}. In spherical polar coordinates $(r, \theta, \varphi)$ this necessarily has $\varphi$-rotational symmetry about the $z$ axis. 

The number of particles $N =  \int_S \rho(\underline{r}) \,dV$ constrains the density, here assumed homogenous $\rho(\underline{r}) = \rho$. If we transform to the frame in which $P$ as the origin, we can write:  

\begin{equation} 
N = \rho \int_S \tilde{r}^2 \, d\tilde{r} \, d\Omega \label{eq:N_Sint}
\end{equation}
where $\tilde{r}$ is the radial component of a point in this frame and $d\Omega$ is the solid angle. Therefore,
\begin{equation} 
\frac{dN}{d\Omega} =  \frac{\rho L(\theta)^3}{3} \coloneqq I(\theta) \label{eq:dN_dOmega}
\end{equation} 

which is the particle mass per solid angle, where $L(\theta)$ is the distance from $P$ to the sphere surface. This quantity $I(\theta)$ is biologically accessible (i.e. can be sensed) via the visual field of an individual within the flock and is closely related (via a threshold function) to the fraction of sky occluded by individuals in the $\theta$ direction as observed from $P$. 

For an individual at $P$ there are intuitively directions which have higher and lower particle mass per solid angle. The imprint of the flock on an individuals visual field is greater when looking through its centre than in the opposite direction, as can be seen in figure \ref{fig:sphere_swarm_schematic}, panels A \& B.

We are interested in the extrema of $I(\theta)$ and make use of the observation that $L(\theta)$ is the radial distance to the flock edge, see figure \ref{fig:sphere_swarm_schematic}, with $P$ as the origin. This has the form $L(\theta) = r_p \cos\theta + \sqrt{R^2 - r_p^2 \sin^2\theta}$. To obtain the extrema of $I(\theta)$ we differentiate equation \ref{eq:dN_dOmega} which yields:
	
\begin{equation} 
\frac{dI}{d\theta} =  - \rho L^2 r_p \sin \theta \Bigg( 1 + \frac{r_p \cos \theta}{\sqrt{R^2 - r_p^2 \sin^2 \theta}} \Bigg) = 0 \label{eq:dI_dTheta}
\end{equation} 

For non-zero density $\rho$, there are a number of stationary points. First when $r_p = 0$, from the perspective of an individual at the centre of the spherical flock, there is no variation in mass density in any direction, and $I(\theta) = \frac{\rho R^3}{3}$ is independent of $\theta$. More significantly, there is a maximum and minimum at $\theta = 0 $ and $\pi$ respectively. These correspond to $L(0) = R + r_p$, looking along a line from $P$ through the centre of the sphere (along $+z$), and $L(\pi) = R - r_p$, away from it (along $-z$). This also provides two features identifiable in the visual field of the individual at $P$: $I_{\max} = \frac{\rho (R + r_p)^3}{3}$ and $I_{\min} = \frac{\rho (R - r_p)^3}{3}$. 

To obtain a quantity which captures the asymmetry of any individual's visual field we take the ratio of the values of these two features to define the ``visual anisotropy'' as:

\begin{equation} 
\Delta I =  \frac{I_{\min}}{I_{\max}} = \frac{(1 - D)^3}{(1 + D)^3} \label{eq:DeltaI}
\end{equation} 

where $D = r_p/R$ is the relative depth within the flock for an individual at $P$. Note how equation \ref{eq:dN_dOmega} does not explicitly feature $\rho$ and is ``scale-free'' by nature, being only a function of the dimensionless depth $D$, and is also monotonic on the interval $D \in [0, 1]$.

\begin{figure} []
	\centering
	\hspace{-3mm}\includegraphics[width=.5\textwidth]{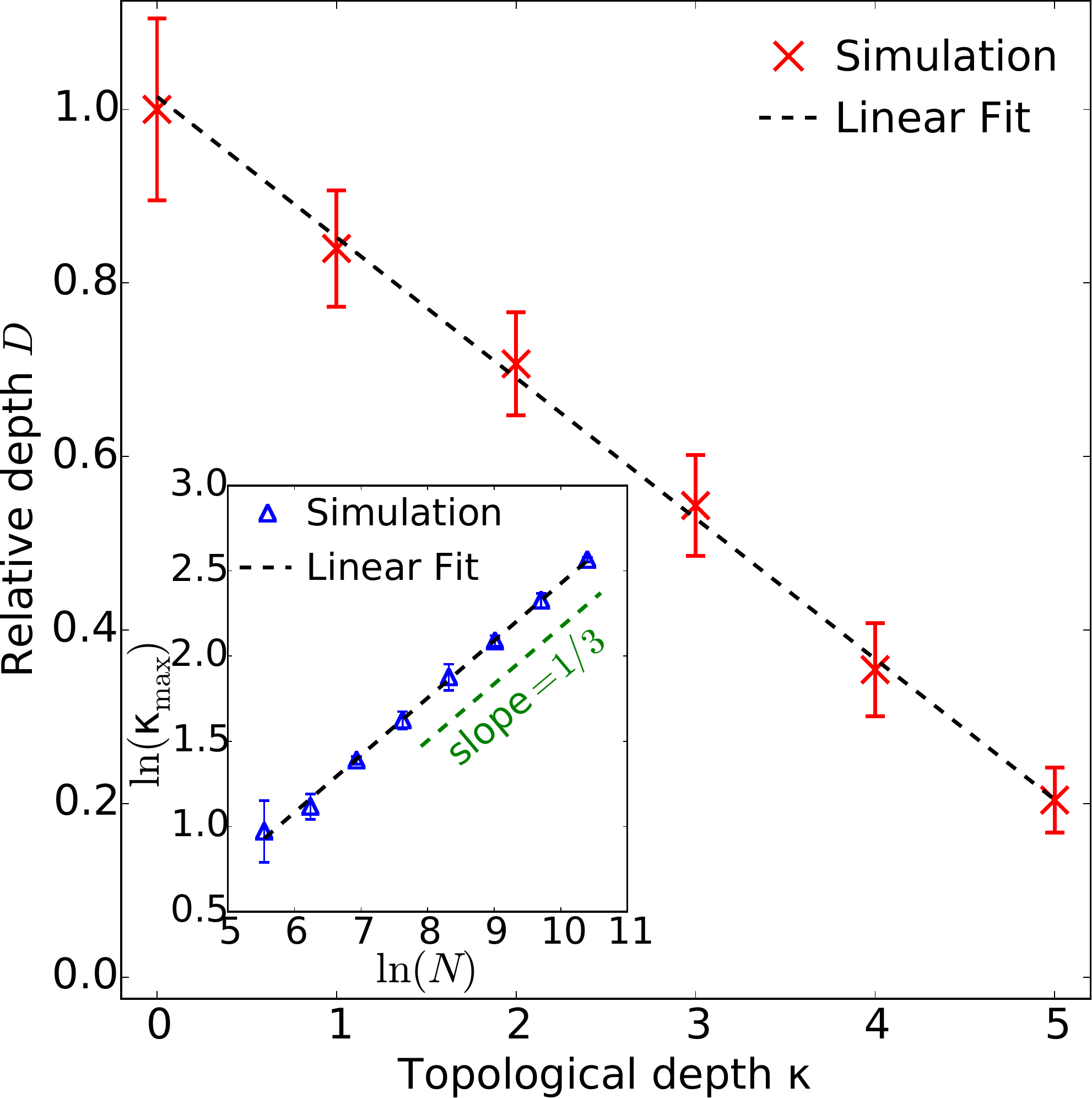}
	\caption{Relative depth $D$ of an individual within a simulated flock compared to its topological depth $\kappa$ averaged over five runs of 1000 time-steps after equilibration with $N = 1500$, $\phi_n = 0.22$ and bounding function as fit to empirical data (figure \ref{fig:opt_bounding_func}) shown as red crosses. Inverse squared-error weighted least squares fit shown as dotted black line: $ D = -0.162 \kappa + 1.015 $. Inset: how maximum topological depth $\kappa_{\max}$ scales with the number of flock members $N$ for the test case of a homogeneous sphere of unit density (blue triangles). Fit function (black dashed line) has the form $\ln(\kappa_{max}) = 0.336 \ln(N) - 0.933$ which suggests $\kappa_{\max} \sim N^{1/3} \sim R$, as one might expect at fixed density. The dashed green line has gradient $1/3$, for reference. Thus maximum topological depth grows with the size of the flock.}
	\label{fig:relative_depth_vs_topo_depth}
\end{figure}

This is useful if it can be linked to topological depth. To make this connection we now seek a relationship between relative depth $D$ and topological depth $\kappa$. For each time-step for our simulated, non-spherical flocks (example configurations can be seen in the Supplemental Materials), we determine the spatial extent of the flock as $R = \langle | \underline{r}_i - \underline{r}_{cm} | \rangle_{i \in C_0}$ the mean distance to centre of mass $\underline{r}_{cm}$ over all particles on the convex hull of the point set. Relative depth per individual is then determined as $D_i = | \underline{r}_i - \underline{r}_{cm} | / R$ which, on average, is one for individuals with zero topological depth. Figure \ref{fig:relative_depth_vs_topo_depth} shows relative depth averaged over a thousand configurations from five simulations with parameters determined from the fit to empirical data, as shown in figure \ref{fig:opt_bounding_func}, compared with the corresponding topological depth. We observe a linear relationship with relative depth decreasing with increased topological depth: when an individual is closer to the centre ($| \underline{r}_i - \underline{r}_{cm} |$ is smaller) it has a higher topological depth and vice versa.

\begin{figure} []
	\centering
 	\includegraphics[width=.48\textwidth]{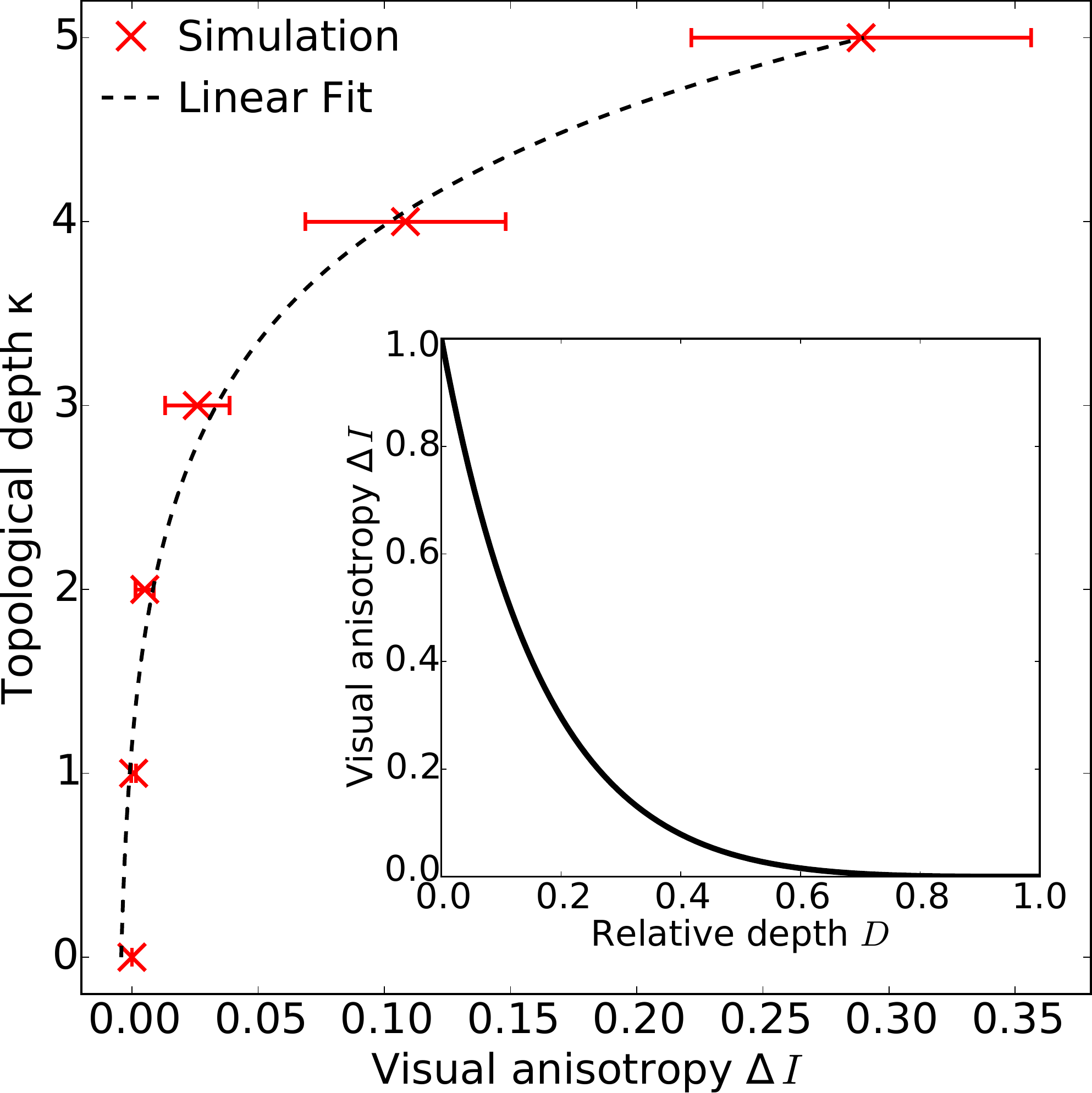}
	\caption{Relationship between visual anisotropy $\Delta I$ and topological depth $\kappa$. Simulation data (red crosses) is as for figure \ref{fig:relative_depth_vs_topo_depth} and unweighted fit (black dashed line) of form $\kappa = a e^{b \Delta I} + c$ is provided as a guide-line ($a = 0.002$, $b = 1.005$, $c = -0.006$). The functional form is not itself important but is designed to show that a simple heuristic relationship could be accessible to animals. Inset: analytic relationship between $\Delta I$ and relative depth $D$ as shown in equation \ref{eq:DeltaI}. Data points from figure \ref{fig:relative_depth_vs_topo_depth} (main) are transformed by this function to obtain the relationship seen in the main figure.}
	\label{fig:topo_depth_vs_deltaI}
\end{figure}

We can finally relate our biologically accessible quantity, the visual anisotropy $\Delta I$, to topological depth $\kappa$ and we show this for our model in Figure \ref{fig:topo_depth_vs_deltaI}, providing a one-to-one map. An individual can therefore compare two features (the minimum and maximum projected density) from their visual field in order to determine their topological depth within the flock, and hence understand how they should adjust their motion. One could imagine such a relationship might be determined heuristically: an intuitive understanding of depth within the aggregation from visual observations. When $\Delta I$ is small, the ratio between minimum and maximum of particle mass per solid angle $I(\theta)$ is large, so there is a large distinction between the two directions these represent (away from and toward the bulk of the flock respectively). When $\Delta I$ is larger the curve has less extreme slope and presents distinct values for different topological depths suggesting an individual deep in the flock still has capacity to determine its depth. We intend to further develop this model, including the role of heterogeneity, in future work \cite{LT_unpublished}.

\section{Conclusion} \label{sec:conclusion}

In conclusion, we have introduced a generalised topological model of collective behaviour with a tunable bounding function to distribute a metric-free motional bias across the swarm. This model was fitted to empirical data of Starling murmurations using stochastic optimisation to determine a suitable form of bounding function. Simulation data from this model was shown to match field study data and produce swarms which are more dense at the border than at the centre, which is a surprising characteristic of real-world Starling flocks. We compared this fitted model to a benchmark topological model with no motional bias on the bulk of the flock (only on the surface). This allowed us to understand the role of the specific form of distributed motional bias that we have identified, which is to produce the desired level of inter-individual exclusion across the swarm, and allow individuals to keep the necessary relative distance apart without directly enforcing what this should be. We also proposed how an individual might use the observed anisotropy of its visual field to determine its depth within the flock.

Models of swarming generally aim to obtain group cohesion and coalignment \cite{reynolds1987flocks,hemelrijk2008self}. Typically, these are explicitly included as rules imposed on the interacting agents in the system. Our model differs from current models in the literature. While it explicitly imposes coalignment in a familiar way, swarm cohesion (and density regulation) are controlled using a motional bias distributed across the flock, which is prescribed via metric-free interaction rules, consistent with experimental observations. We show that specific field observations of density variation in aggregations of Starlings can be reproduced using our model so that density is higher on the border of the flock than at the centre. This density profile may relate to the predator-evasion mechanisms of three-dimensional swarms and the evolutionary development of such behaviour.

\ack
This work was funded by the UK Engineering and Physical Sciences Research Council (EPSRC) through the Complexity Science Doctoral Training Centre (JML), Grant EP/E501311 and by EPSRC grant \# EP/E501311/1, a Leadership fellowship (MST). Computing facilities were provided by the Centre for Scientific Computing of the University of Warwick with support from the Science Research Investment Fund. We also acknowledge the use of the Computational Geometry Algorithms Library (CGAL) \cite{cgal}. The authors would like to thank the referees for encouraging us to think more about how topological depth might be sensed.

\section*{References}
\bibliographystyle{iopart-num}
\bibliography{LewisTurner_DMBSMF_JPhysD_AcceptedManuscript}

\end{document}